\documentclass[aps,prb,twocolumn,floatfix,showpacs,superscriptaddress]{revtex4-2}
\usepackage{graphics}
\usepackage{epsfig}
\usepackage{times}
\usepackage{bm}
\usepackage{braket}
\usepackage{color}

\usepackage{amsmath,amssymb}	
\begin{document}
\title{Competing Magnetic and Topological Orders in the Spin-1 Kitaev-Heisenberg Chain\\ with Single-Ion Anisotropy}
\author{Sahinur Reja}
\affiliation{Department of Physics, Jadavpur University, Kolkata 700032,
West Bengal, India}
\author{Satoshi Nishimoto}
\affiliation{Department of Physics, Technical University Dresden, 01069 Dresden, Germany}
\affiliation{Institute for Theoretical Solid State Physics, IFW Dresden, 01069 Dresden, Germany}

\date{\today}

\begin{abstract}
	We investigate the ground-state phase diagram of the spin-1 Kitaev--Heisenberg chain in the presence of uniaxial single-ion anisotropy (SIA) $D_z$ by density-matrix renormalization group (DMRG) calculations. By combining energy-curvature diagnostics on periodic $N=24$ clusters with a refined characterization based on order parameters and correlation functions for open chains up to $N=144$, we establish a comprehensive phase diagram in the $\phi$--$D_z$ plane. We identify four magnetically ordered phases---FM-$z$, FM-$xy$, N\'eel-$z$, and a two-sublattice collinear LLRR2 state---as well as magnetically disordered/critical regimes including N\'eel-$xy$, LLRR1, and two Kitaev spin-liquid (KSL) regions. A topological Haldane phase also emerges near the Heisenberg limit. Our results provide evidence that both AFM- and FM-KSL regimes acquire finite parameter widths in the spin-1 model, while the Haldane phase is fragile against Kitaev-type anisotropy, particularly for $D_z<0$. Increasing (decreasing) $D_z$ suppresses (enhances) magnetic order and expands (shrinks) the KSL and other magnetically disordered sectors. Also, at $D_z=0$, we identify an exactly solvable point at $\phi=\tan^{-1}(-2)$, which enforces a first-order transition between N\'eel-$z$ and LLRR2. We further contrast these findings with the spin-$1/2$ KH chain and with the spin-1 honeycomb KH model, highlighting the distinct roles of dimensionality and SIA in Kitaev-type magnets.
\end{abstract}

\maketitle

\section{introduction}

Frustrated quantum magnets in low dimensions provide a fertile arena for unconventional phases and exotic excitations, including quantum spin liquids (QSLs) featuring fractionalization and topological order~\cite{Balents2010,Savary2017,Zhou2017,Knolle2019}. Among the theoretical platforms, the Kitaev model with bond-dependent anisotropic exchange has played a central role in elucidating such physics~\cite{Kitaev2006}. In realistic materials, however, conventional Heisenberg interactions are typically present as well, motivating extensive studies of the Kitaev--Heisenberg (KH) model in one and two dimensions. The resulting phase diagrams host highly entangled ground states with fractionalized excitations and topological order in certain parameter regimes, and are widely discussed in connection with topological quantum computation and emergent gauge structures~\cite{Anderson1973,Wen1991,Aasen2020}.

While most prior work has focused on spin-1/2 systems~\cite{Winter2017,Trebst2022}, recent experimental progress has broadened the search to higher-spin magnets, including spin-1 candidates such as Na$_2$Ni$_2$TeO$_6$~\cite{Kurbakov2020,Samarakoon2021,Bera2022}, A$_3$Ni$_2$SbO$_6$ ($A$=Li, Na)~\cite{Zvereva2015,Kurbakov2017,Nalbandyan2025}, Na$_3$Ni$_2$BiO$_6$~\cite{Shangguan2023}, and KNiAsO$_4$~\cite{Taddei2023,Haraguchi2025}. For $S>1/2$, the physics is richer and more challenging. In contrast to the $S=1/2$ case, the pure $S=1$ Kitaev model is not exactly solvable, and no exact solution in terms of free Majorana fermions coupled to a static ${\cal Z}_2$ gauge field is known. Accordingly, a variety of numerical and analytical approaches have been employed to clarify the fundamental properties of the spin-1 Kitaev model~\cite{Koga2018,Oitmaa2018,Dong2020,Zhu2020,Hickey2020,Lee2020,Khait2021,Bradley2022,Fukui2022,Pohle2023,Consoli2020,Chen2022,Chen2023,Mashiko2024,Ralko2024,Ayushi2024,Luo2024,Georgiou2024,Sasamoto2025}. Notably, compared to the spin-1/2 case where the Kitaev spin-liquid (KSL) regime can be broad~\cite{Chaloupka2010}, spin-1 systems typically display enhanced stability of magnetically ordered phases due to reduced quantum fluctuations~\cite{Fukui2022}.

An efficient route toward understanding the ground-state properties of two-dimensional (2D) KH systems is to study their one-dimensional (1D) analogs, since many ordered states in the 2D spin-1/2 KH model can be rationalized in terms of the corresponding 1D KH chain~\cite{Clio2018,Yang2020,Yang2025}, which allows substantially more accurate numerical analysis. Along these lines, several theoretical works have investigated the spin-1 KH chain~\cite{You2020,You2022,Zhang2023,Zhang2024}. Spin-1 chains additionally support the topological Haldane phase protected by hidden symmetry~\cite{Haldane_1,AKLT_PhysRevLett.59.799}. The overall phenomenology resembles that of the spin-1/2 KH chain, except that the Tomonaga--Luttinger liquid (TLL) near the Heisenberg limit is replaced by the Haldane phase and the KSL regimes are suggested to acquire finite parameter widths rather than remaining isolated points~\cite{You2020}.

Another important ingredient specific to $S>1/2$ systems is uniaxial single-ion anisotropy (SIA). In Mott insulators, onsite interactions can generate nonlinear spin terms of the form $D_z\sum_i(S_i^z)^2$, where $D_z$ controls easy-plane ($D_z>0$) versus easy-axis ($D_z<0$) tendencies. Recent studies of spin-1 Kitaev materials incorporating SIA have attracted increasing attention~\cite{Bradley2022,Luo2023,Ayushi2024}. In general, positive $D_z$ suppresses magnetic order and tends to broaden KSL and/or nonmagnetic sectors, whereas negative $D_z$ stabilizes Ising-like N\'eel-$z$ and ferromagnetic (FM)-$z$ orders. In 2D systems, SIA can also promote more exotic textures; for instance, in the spin-1 honeycomb KH model, large positive $D_z$ favors vortex states with enlarged magnetic unit cells that erode the KSL region~\cite{Ayushi2024}. Motivated by its clear relevance to spin-1 Kitaev candidates~\cite{Zvereva2015,Kurbakov2017,Nalbandyan2025,Taddei2023,Haraguchi2025}, where SIA is often unavoidable, the combined effect of Kitaev, Heisenberg, and SIA terms in spin-1 chains remains comparatively unexplored.

In this paper, we accurately determine the ground-state phase diagram of the spin-1 KH chain with SIA by density-matrix renormalization group (DMRG) calculations. We express the Heisenberg and Kitaev exchanges with a single parameter $\phi$ as $J=\cos\phi$ and $K=\sin\phi$ and map out the $\phi$--$D_z$ plane by first locating putative transitions from the second derivative of the ground-state energy on $N=24$ periodic chains. We then diagnose the phases via relevant order parameters and correlation functions for longer open chains up to $N=144$ and determine accurate phase boundaries by thermodynamic extrapolations of these order parameters. Our calculations reveal four magnetically long-range-ordered (LRO) phases [FM-$z$, FM-$xy$, a two-sublattice collinear ($\mathrm{LLRR2}$) state, and N\'eel-$z$], two magnetically short-range-ordered (SRO) regions [N\'eel-$xy$ and $\mathrm{LLRR1}$], and a topological Haldane phase, together with antiferromagnetic (AFM) and FM KSL regimes, hereafter referred to as AFM-KSL and FM-KSL, which acquire \emph{finite} parameter widths in the spin-1 case. We further identify an exactly solvable point at $\phi=\tan^{-1}(-2)$ for $D_z=0$, which enforces a first-order boundary between N\'eel-$z$ and $\mathrm{LLRR2}$.

The rest of the paper is organized as follows. In Sec.~\ref{sec:model}, we describe the model. In Sec.~\ref{sec:method}, we provide details of the numerical techniques and define the order parameters used to identify the phases and their boundaries. In Sec.~\ref{sec:results}, we present the numerical results leading to the $\phi$--$D_z$ phase diagram. Finally, we provide discussion and summary in Secs.~\ref{sec:discussion} and \ref{sec:summary}, respectively.

\section{Model}\label{sec:model}

\begin{figure}[tbh]
	\centering
	\includegraphics[width=0.9\linewidth]{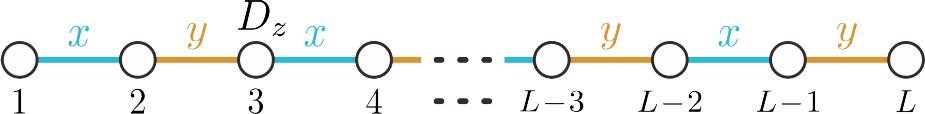}
	\caption{
		Lattice structure of the spin-1 KH chain with SIA. The chain consists of alternating $x$ and $y$ bonds, indicated by the labels `x' and `y' (see text). This alternation mimics the bond-directional pattern of the Kitaev interaction on the honeycomb lattice projected onto one dimension.
	}
	\label{fig:lattice}
\end{figure}

We consider the spin-1 KH chain with SIA, described by the Hamiltonian
\begin{align}
	\mathcal{H} = \sum_{i} \left[ J\, \mathbf{S}_i \cdot \mathbf{S}_{i+1} + K\, S_i^\gamma S_{i+1}^\gamma + D_z(S_i^z)^2 \right],
	\label{eq:ham}
\end{align}
where $\mathbf{S}_i = (S_i^x, S_i^y, S_i^z)$ is the spin-1 operator on site $i$, and $\gamma = x$ or $y$ depending on the bond type (see Fig.~\ref{fig:lattice}). In our 1D setup, we impose a two-site unit cell, where the bond-dependent Kitaev interactions alternate between $x$ and $y$ bonds along the chain. Explicitly, the bond type $\gamma$ is assigned as
\begin{align}
	\gamma =
	\begin{cases}
		x & \text{if } i \text{ is odd}, \\
		y & \text{if } i \text{ is even},
	\end{cases}
	\label{eq:xy}
\end{align}
so that the chain starts with an $x$ bond at the left edge under open boundary conditions (OBC). We define the order parameters corresponding to various magnetic phases according to this convention (see below). 
The alternation of $x$ and $y$ bonds reproduces the bond-directional structure of the Kitaev exchange on the honeycomb lattice projected onto a 1D geometry.

The parameters $J$ and $K$ denote the isotropic Heisenberg and anisotropic Kitaev exchange couplings, respectively. The SIA $D_z$ controls the energetic cost of nonzero $S^z$ components, favoring easy-plane ($D_z > 0$) or easy-axis ($D_z < 0$) behavior depending on its sign. To comprehensively explore the ground-state properties of our model \eqref{eq:ham} as a function of the exchange parameters $J$ and $K$, we parametrize them as
\begin{align}
	J = \cos\phi, \quad K = \sin\phi,
	\label{eq:para_phidep}
\end{align}
so that the model can be analyzed over the full circle $\phi \in [0, 2\pi)$. 
This parametrization allows us to smoothly interpolate between various limiting cases, including the pure Heisenberg ($K = 0$) and pure Kitaev ($J = 0$) chains, as well as mixed regimes where the two interactions compete. Unless otherwise stated, we set the overall exchange energy scale to $\sqrt{J^2 + K^2} = 1$. 




\section{Method}\label{sec:method}

In this section, we summarize our DMRG setup and explain how we combine periodic- and open-boundary calculations to construct the phase diagram and to identify each phase.

\subsection{Numerical technique}

To investigate the ground-state properties of the model \eqref{eq:ham}, we employ the DMRG method~\cite{White1992}, which provides high accuracy for 1D systems with short-range interactions. We use both periodic boundary conditions (PBC) and open boundary conditions (OBC); their roles are complementary, as described below.

For bulk quantities such as the ground-state energy $E_0$, the total spin $S_{\rm tot}$, and the static spin structure factor $S(q)$, we perform PBC calculations on $N=24$-site rings. Since PBC calculations are more demanding, we retain up to $m=4000$ density-matrix eigenstates and perform up to 20 sweeps to ensure convergence. The typical discarded weight (truncation error) is less than $10^{-8}$.

For nonlocal probes (e.g., string correlations) and for evaluating order parameters corresponding to various magnetic phases, we perform OBC calculations, which yield higher accuracy at comparable computational cost. In OBC calculations, we may employ weak boundary pinning fields or symmetry-broken initial states to select a representative ordered configuration, thereby enabling the direct identification of magnetic order through local expectation values (e.g., $\langle S_i^z\rangle$) in the bulk. We study OBC chains with $N=24--144$ sites, retain up to $m=4000$ states, and achieve truncation errors $\lesssim 10^{-10}$. To minimize boundary-induced Friedel oscillations, OBC observables are evaluated in a central window of sites, far from the edges. This choice reduces finite-size corrections and improves the stability of the subsequent finite-size scaling analyses. Order parameters obtained for various system sizes are extrapolated to the thermodynamic limit. Unless stated otherwise, all results are converged with respect to $m$, the number of sweeps, and system sizes.

\subsection{Investigation procedure}

We first obtain a global characterization of the ground-state phases by systematically varying the coupling angle $\phi$ [Eq.~(3)] and scanning the SIA strength $D_z$. The phase diagram is initially mapped on a small PBC cluster ($N=24$ site ring) by evaluating the following bulk diagnostics: (i) the ground-state energy per site $E_0/N$, (ii) the total spin per site
$S_{\rm tot}/N=\langle \mathbf{S}_{\rm tot}^2\rangle/N$, and (iii) the static spin structure factor
\begin{equation}
	S^{\alpha\alpha}(q)=\frac{1}{N}\sum_{i,j}e^{iq(r_i-r_j)}\langle S_i^\alpha S_j^\alpha\rangle
	\qquad (\alpha=x,y,z),
	\label{eq:Sq}
\end{equation}
whose peak position diagnoses magnetic periodicity via a characteristic wave vector $q$. This strategy provides a reliable starting point for locating magnetic phase boundaries in related KH models~\cite{Chaloupka2013,Agrapidis2018,Kadosawa2023}.

Once a coarse phase diagram is obtained, we refine the phase boundaries and determine the nature of each phase (LRO vs. SRO) by performing higher-accuracy OBC calculations on longer chains. Specifically, we compute real-space correlations $\langle S_i^\alpha S_{i+r}^\alpha\rangle$ and evaluate order parameters that distinguish topological order, magnetic order, and symmetry breaking. The order parameters used to identify the phases summarized in Fig.~\ref{fig:states} are defined below.

\subsection{Order parameters}

\begin{figure}[t]
	\centering
	\includegraphics[width=1.0\linewidth]{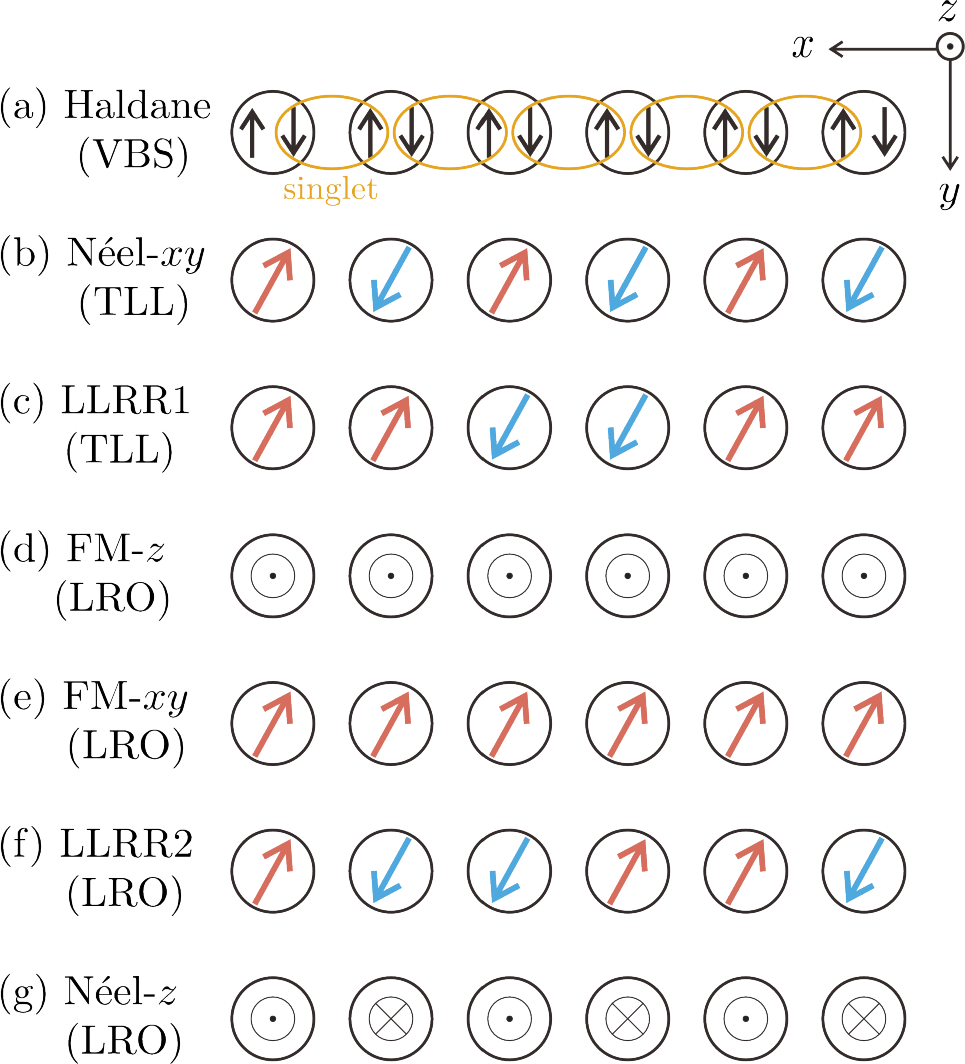}
	\caption{Schematic representations of the seven phases realized in the 1D spin-1 KH model, excluding the KSL regimes. We do not include the KSL states here because they do not exhibit a simple semiclassical spin pattern associated with local symmetry breaking. In the ideal Kitaev limit, spin correlations are extremely short-ranged (only nearest-neighbor bond correlations remain finite), which further complicates a schematic real-space depiction.}
	\label{fig:states}
\end{figure}

To characterize the ground-state phases of the Hamiltonian in Eq.~\eqref{eq:ham}, we introduce appropriate order parameters for the relevant candidate phases. Phases with spontaneous symmetry breaking are identified by conventional magnetic order parameters, whereas topological and disordered phases require nonlocal quantities or bond-energy–based observables. Within DMRG simulations on open chains, these order parameters are constructed from local spin expectation values and correlation functions, 
taking advantage of the boundary conditions. In particular, open edges explicitly break both spin-rotational and translational symmetries, thereby selecting a preferred ordering direction. This controlled symmetry breaking enables accurate evaluation of local order parameters and reliable finite-size scaling analyses. In the following, we define the specific order parameters used to distinguish the phases summarized in Fig.~\ref{fig:states}.

\subsubsection{Haldane phase}

The Haldane phase does not exhibit conventional magnetic order, but it can be characterized as a topological valence bond solid (VBS) state with hidden $\mathbb{Z}_2 \times \mathbb{Z}_2$ symmetry breaking~\cite{den_Nijs1989,Kennedy1992,Oshikawa1992}. This phase is typically identified by evaluating the nonlocal string correlation function
\begin{align}
\mathcal{C}_\mathrm{str}(i,j)= \left\langle S^z_i \exp\left( i\pi \sum_{k=i+1}^{j-1} S^z_k \right) S^z_j \right\rangle,
\label{eq:string}
\end{align}
whose asymptotic value $|\mathcal{C}_\mathrm{str}(-\infty,\infty)|$ serves as a string order parameter that characterizes the Haldane phase. In practical simulations, we define the string order parameter around the center of the system as
\begin{align}
	{\cal O}_{\rm Haldane}=\lim_{N\to\infty}\left|\mathcal{C}_\mathrm{str}\left(\frac{N}{4},\frac{3N}{4}\right)\right|.
	\label{eq:haldane}
\end{align}
For the spin-1 Heisenberg chain ($\phi=0$), this parameter takes the value $|\mathcal{C}_\mathrm{str}(-\infty,\infty)| \approx 0.3743$~\cite{White1993}.

\subsubsection{N\'eel-$xy$ phase}

The N\'eel-$xy$ phase is characterized by alternating spin alignment in the $xy$-plane. We define the corresponding order parameter from the local spin expectation values at the chain center as
\begin{align}
	\mathcal{O}_{\rm N\acute{e}el-xy} 
	= \lim_{N \to \infty} 
	\left| \langle S^x_{N/2} \rangle - \langle S^x_{N/2+1} \rangle \right|.
	\label{eq:neelxy}
\end{align}
To induce symmetry breaking and stabilize a preferred orientation, staggered boundary fields are applied along the $x$ axis: $h^x_1 = -h^x_N = h$.  
Although strong edge fields polarize spins near the boundaries, the bulk value of $\mathcal{O}_{\rm N\acute{e}el-xy}$ converges to a well-defined limit independent of $h$ as $N \to \infty$, allowing reliable extraction from the central region.

Because of remaining spin-rotational symmetry in the $xy$-plane, an equivalent definition can be given using the $y$-component,
\begin{align}
	\mathcal{O}_{\rm N\acute{e}el-xy} 
	= \lim_{N \to \infty} 
	\left| \langle S^y_{N/2} \rangle - \langle S^y_{N/2+1} \rangle \right|,
\end{align}
with $h^y_1 = -h^y_N = h$. In practice, the open boundaries spontaneously select one direction, so either choice is valid in numerical simulations.

\subsubsection{N\'eel-$z$ phase}

The N\'eel-$z$ phase is defined analogously, but the spin symmetry is broken only along the $z$ axis. The corresponding order parameter is uniquely given by
\begin{align}
	\mathcal{O}_{\rm N\acute{e}el-z} 
	= \lim_{N \to \infty} 
	\left| \langle S^z_{N/2} \rangle - \langle S^z_{N/2+1} \rangle \right|,
	\label{eq:neelz}
\end{align}
where staggered boundary fields $h^z_1 = -h^z_N = h$ are applied to stabilize the symmetry-broken state.

\subsubsection{FM-$xy$ phase}

The FM-$xy$ phase is characterized by uniform FM alignment in the $xy$-plane. Assuming that the spontaneous magnetization points along the $x$ axis, the order parameter is defined from the bulk magnetization as
\begin{align}
	\mathcal{O}_{\rm FM-xy} 
	= \lim_{N \to \infty} \left| \langle S^x_{N/2} \rangle \right|.
	\label{eq:fmxy}
\end{align}
To stabilize the symmetry-broken state, uniform boundary fields are applied along $x$: $h^x_1 = h^x_N = h$.

\subsubsection{FM-$z$ phase}

The FM-$z$ phase exhibits FM order along the $z$ axis. The corresponding order parameter is given by
\begin{align}
	\mathcal{O}_{\rm FM-z} 
	= \lim_{N \to \infty} \left| \langle S^z_{N/2} \rangle \right|,
	\label{eq:fmz}
\end{align}
with uniform boundary fields $h^z_1 = h^z_N = h$ applied to select the ordered state.

\subsubsection{LLRR1 phase}

The LLRR1 phase occurs in the regime $-J \ll K \approx 1$ ($J<0$) and preserves spin-rotational symmetry in the $xy$ plane. Assuming the magnetization axis along $x$, the dominant AFM interaction resides on $x$ bonds, while $y$ bonds are weaker and FM, giving rise to a four-site modulation. For open chains with $N=4n$, we define the order parameter using four central sites:
{\small
\begin{align}
	\mathcal{O}_{\rm LLRR1} 
	= \lim_{N \to \infty} \tfrac{1}{4} 
	\left| \langle S^x_{L/2-1} \rangle - \langle S^x_{L/2} \rangle 
	+ \langle S^x_{L/2+1} \rangle - \langle S^x_{L/2+2} \rangle \right|.
	\label{eq:llrr1}
\end{align}
}
Boundary fields are applied as $h^x_1 = h^x_N = h$.

\subsubsection{LLRR2 phases}

The LLRR2 phase has a similar spin pattern to LLRR1, also with four-site periodicity, but it emerges in the regime $J \ll -K \approx 1$ ($J>0$). Again, assuming the magnetization axis along $x$, the $x$-bond interaction is dominant and FM, while the $y$-bond is weaker and AFM. The order parameter can be defined as:
{\small
\begin{align}
	\mathcal{O}_{\rm LLRR2} = \lim_{N\to\infty} \frac{1}{4} \left| \langle S^x_{L/2 - 1} \rangle + \langle S^x_{L/2} \rangle - \langle S^x_{L/2 + 1} \rangle - \langle S^x_{L/2 + 2} \rangle \right|.
	\label{eq:llrr2}
\end{align}
}
This is similar in form to the LLRR1 order parameter but shifted by a relative phase of $\pi/2$. The boundary fields are applied as $h^x_1 = -h^x_N = h$.

\subsubsection{Kitaev spin-liquid phase}

The KSL phase does not exhibit any magnetic long-range order. Still, it can be identified by the absence of conventional order parameters and the presence of short-range entanglement-the basic characteristic of quantum spin liquids. In our model, we detect the onset of dimerization associated with bond-dependent Kitaev interactions by introducing a dimer order parameter
\begin{align}
	\mathcal{O}_{\rm dimer}^\gamma = \lim_{N\to\infty} \left| \langle S_i^\gamma S_{i+1}^\gamma \rangle - \langle S_{i+1}^\gamma S_{i+2}^\gamma \rangle \right|,
	\label{eq:dimer}
\end{align}
where $\gamma = x, y, z$ indicates the bond type and $i$ is chosen near the center of the chain. In the KSL regime, this quantity remains small and size-independent, reflecting the absence of spontaneous dimer formation. A finite value of this order parameter in the thermodynamic limit signals a dimerized phase, while a vanishing value indicates a spin-liquid-like disordered phase.

\section{Numerical Results}\label{sec:results}

\subsection{Second derivative of the energy}

\begin{figure}[t]
	\centering
	\includegraphics[width=1.0\linewidth]{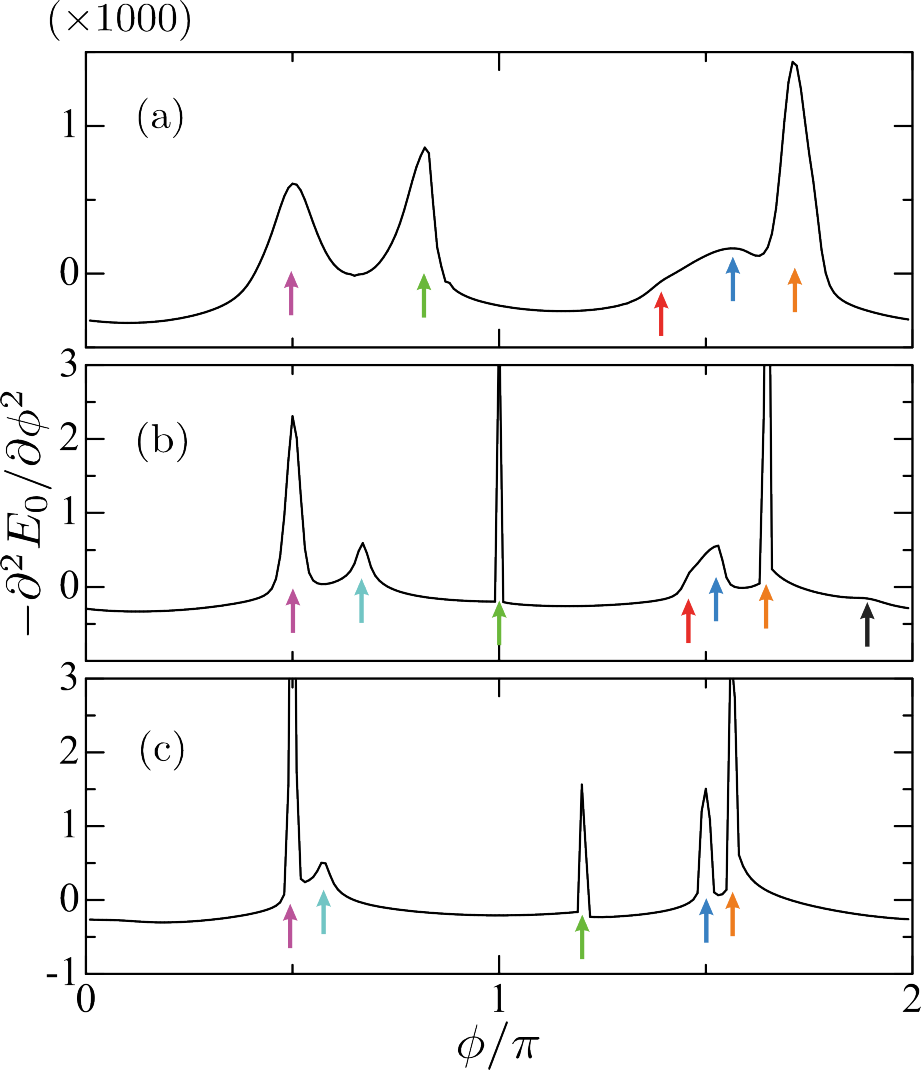}
	\caption{Second derivative of the ground-state energy $E_0$ with respect to $\phi$ for fixed (a) $D_z=0.6$, (b) $0.0$, and (c) $-0.6$. The arrows indicate possible phase boundaries.}
	\label{fig:2ndDE}
\end{figure}

\begin{figure}[thb]
	\centering
	\includegraphics[width=1.0\linewidth]{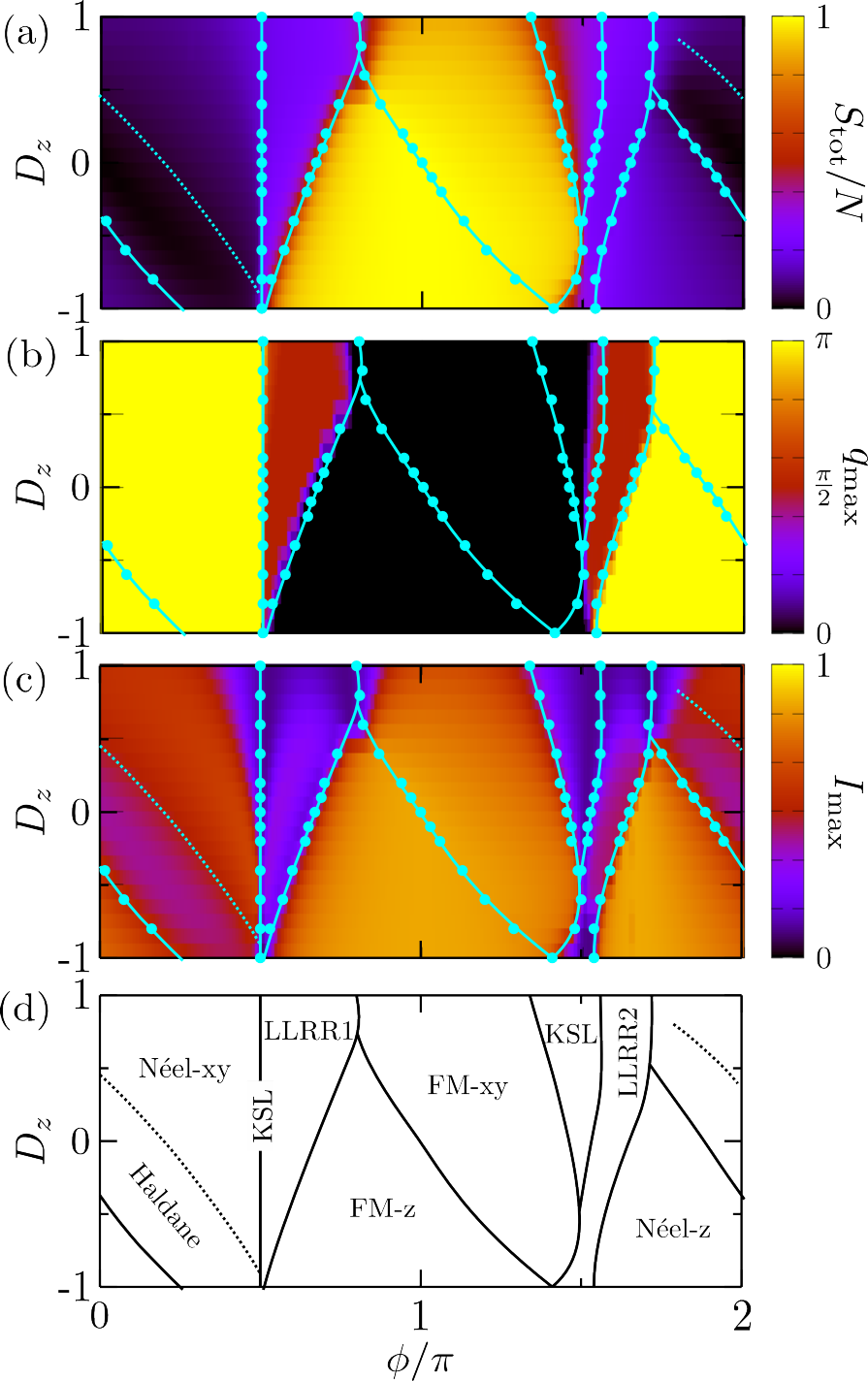}
	\caption{Color maps of (a--c) the total spin $S_{\mathrm{tot}}$, the momentum $q_{\mathrm{max}}$ at which the spin structure factor $S^{\alpha\alpha}(q)$ attains its maximum, and the corresponding peak intensity $I_{\mathrm{max}}$, plotted as functions of $\phi/\pi$ and $D_z$. The transition points extracted from Fig.~\ref{fig:2ndDE} are also indicated. (d) Ground-state phase diagram in the $\phi$--$D_z$ plane, obtained from the analysis of the second derivative of the energy. The dotted line is approximately evaluated from $I_{\mathrm{max}}$ (see text).
	}
	\label{fig:totS}
\end{figure}

To establish the basic structure of the phase diagram in $\phi-D_z$ plane, we start by evaluating the second derivative of the ground-state energy $E_0$ with respect to the tuning parameter $\phi$ and for different values of $D_z$, which serves as a sensitive probe for identifying phase transitions. In general, peaks or shoulders in this quantity indicate the locations of phase boundaries. Fig.~\ref{fig:2ndDE}(a--c) displays the second derivative of $E_0$ as a function of $\phi/\pi$ for three representative values of $D_z=0.6, 0.0$ and $-0.6$ respectively, obtained for a periodic chain of $N=24$ sites. The positions of these anomalies---indicated by arrows---are taken as estimates of the transition points.  

To determine which phases are separated by these transitions, we examine the color maps of the total spin $S_{\mathrm{tot}}$, the momentum $q_{\mathrm{max}}$ at which the spin structure factor $S^{\alpha\alpha}(q)$ attains its maximum, and the corresponding peak intensity $I_{\mathrm{max}}$ as depicted respectively in Fig.~\ref{fig:totS}(a--c). These quantities are plotted as functions of $\phi/\pi$ and $D_z$, together with the transition points extracted from Fig.~\ref{fig:2ndDE}. Based on these results, the overall phase diagram as a function of $\phi/\pi$ and $D_z$ is presented in Fig.~\ref{fig:totS}(d). We note that the phase boundary between the Haldane and N\'eel-$xy$ phases does not produce a pronounced peak in the second derivative of the ground-state energy. We therefore estimate this boundary approximately from the behavior of $I_{\mathrm{max}}$: in the Haldane phase, spin--spin correlations decay exponentially with distance, resulting in a comparatively small peak intensity of $S^{\alpha\alpha}(q)$, see Fig.~\ref{fig:totS}(c).

However, this approach has limited sensitivity to transitions between magnetically disordered phases, where conventional order parameters do not exhibit singular behavior. In our model, this limitation is the most evident near the boundaries separating the AFM Kitaev phase from its neighboring N\'eel-xy and LLRR1 phases, as well as at the transition between the Haldane and N\'eel-$xy$ regions as mentioned above. Nevertheless, previous studies have demonstrated that phase diagrams inferred from the second derivative of the ground-state energy, even for moderately large finite systems, provide a reliable approximation to the thermodynamic limit in case of magnetically ordered phases. Motivated by this observation, we use the tentative phase boundaries identified here as a starting point for a more detailed analysis of the order parameters characterizing each phase. These analyses are presented in the following subsections.

\subsection{Phase diagram at $D_z=0$}

\begin{figure}[thb]
	\centering
	\includegraphics[width=1.0\linewidth]{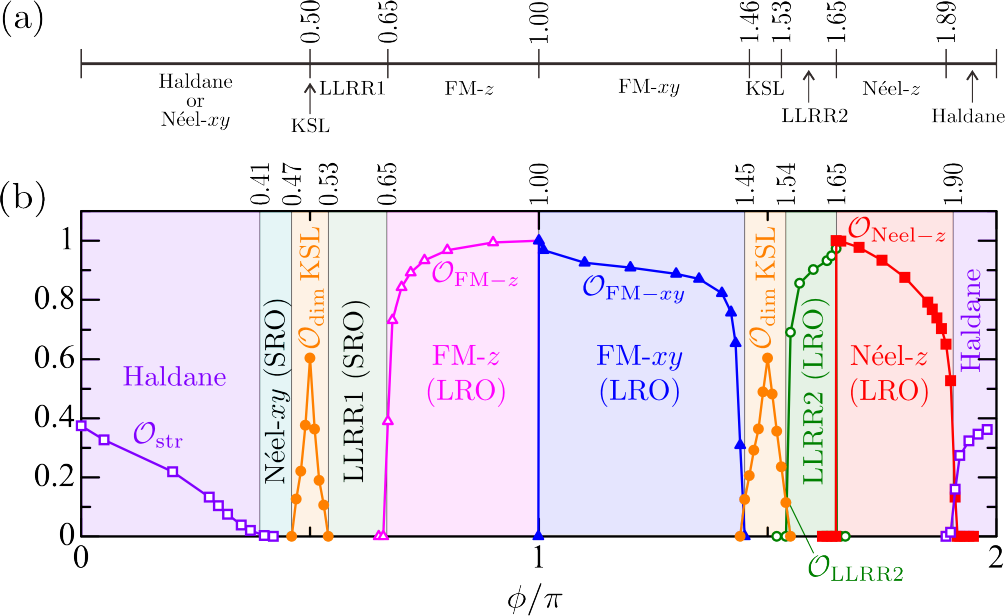}
	\caption{Ground-state phase diagram versus $\phi/\pi$ at $D_z=0$, obtained from (a) the second derivative of the energy for a 24-site PBC cluster and (b) extrapolated order parameters in the thermodynamic limit.}
	\label{fig:Dz0c00}
\end{figure}

\subsubsection{Comparison between the second-derivative and order-parameter approaches}

We first focus on the $D_z=0$ case. As a reference, Fig.~\ref{fig:Dz0c00}(a) shows the phase diagram as a function of $\phi/\pi$ obtained from the above analysis on an $N=24$ PBC cluster. We further compute the relevant order parameters for open chain systems of length up to $N=144$-sites and extrapolate them to the thermodynamic limit; the extrapolated values of the order parameters are summarized in Fig.~\ref{fig:Dz0c00}(b). Details of the finite-size scaling analyses are provided in Appendix~\ref{App:scaling}.

As expected, the phase boundaries between magnetically ordered phases, as well as those between ordered and disordered phases, agree quantitatively well between the two methods. However, for boundaries between magnetically disordered phases, the situation is somewhat different. In particular, the phase diagram obtained from the order parameters reveals an additional boundary between the Haldane and N\'eel-$xy$ phases, which was not captured by the analysis of second-derivative of energy. Moreover, while the second-derivative approach suggested a single transition point near $\phi/\pi = 0.5$, the order-parameter analysis shows that the KSL phase actually occupies a finite parameter region.

\subsubsection{Order-parameter behavior for magnetically ordered phases}

We now turn to a more detailed quantitative examination of each order parameter. We begin with four phases exhibiting magnetic LRO: FM-$z$ ($0.65 \lesssim \phi/\pi < 1.00$), FM-$xy$ ($1.00 \lesssim \phi/\pi < 1.45$), LLRR2 ($1.54 \lesssim \phi/\pi < 1.65$), and N\'eel-$z$ ($1.65 \lesssim \phi/\pi < 1.90$). In each case, the corresponding order parameters---$\mathcal{O}_{\mathrm{FM\text{-}z}}$, $\mathcal{O}_{\mathrm{FM\text{-}xy}}$, $\mathcal{O}_{\mathrm{LLRR2}}$, and $\mathcal{O}_{\mathrm{N\acute{e}el\text{-}z}}$---extrapolate to a finite value in the thermodynamic limit. The FM-$z$ and FM-$xy$ phases, as well as the LLRR2 and N\'eel-$z$ phases, are directly adjacent to each other, and the transitions between them are of first-order character, as evidenced by discontinuous jumps in the order parameters as depicted in Fig.~\ref{fig:Dz0c00}(b). Each of these ordered phases also borders a disordered phase. Near the phase boundaries ($\phi/\pi \approx 0.65, 1.45, 1.54, 1.90$), the order parameters exhibit sharp but continuous increases, suggesting that these transitions are either weakly first order or nearly first order in nature. Since such transitions are typically accompanied by level crossings in finite systems, the transition points identified from the second derivative agree quantitatively with those extracted from the order-parameter analysis [cf.~Fig.~\ref{fig:Dz0c00}(a)].

{\it Ferromagnetic phases.}--To understand the phases better, we utilize the fact that the Hamiltonian in Eq.~\eqref{eq:ham} without SIA term can be written as the sum of Ising, exchange and double-spin-flip terms as $H = {\cal H}_{\text{Ising}} + {\cal H}_{\text{ex}} + {\cal H}_{\text{dsf}}$ with ${\cal H}_{{\rm ex},i} =\frac{2J+K}{4} (S_i^+S_{i+1}^- + S_i^-S_{i+1}^+)$, ${\cal H}_{{\rm dsf},i} = \frac{K}{4}(-1)^i S_i^+S_{i+1}^+ + S_i^-S_{i+1}^-$, and ${\cal H}_{{\rm Ising},i} = J S_i^zS_{i+1}^z$. Since both coefficients of ${\cal H}_{\rm ex}$ and ${\cal H}_{\rm Ising}$ are negative, an FM LRO is naively expected in the range $\tan^{-1}(-2)\approx0.65 \lesssim \phi/\pi \lesssim 1.5$. In fact, we have found an FM phase for $0.65 < \phi/\pi < 1.46$ at $D_z=0$. At the spin-isotropic point $\phi/\pi = 1$, our system is equivalent to the spin-1 FM Heisenberg chain. Thus, the spins can align along any arbitrary spatial direction owing to the SU(2) rotational invariance, and the total spin takes its maximum value, $\mathcal{O}_{\mathrm{FM\text{-}xy}} = \mathcal{O}_{\mathrm{FM\text{-}z}} = 1$, as shown in Fig.~\ref{fig:Dz0c00}(b). Away from the isotropic point, the SU(2) symmetry is broken. For $\phi/\pi < 1$, the magnitude of $|J^z|$ is comparable to or larger than that of $|J^x|$ and $|J^y|$, so the magnetization axis is oriented along the $z$ direction. In contrast, for $\phi/\pi > 1$, the $|J^z|$ component becomes comparable to or smaller than $|J^x|$ and $|J^y|$, leading to a magnetization axis lying within the $xy$ plane. Within this plane, however, the magnetization direction can freely rotate, reflecting the residual U(1) rotational symmetry.

{\it Exactly solvable point.}---At $D_z=0$, in addition to the isotropic ferromagnetic point ($\phi/\pi = 1$), we find that the model in Eq.~\eqref{eq:ham} can also be solved analytically at $\phi = \tan^{-1}(-2) \approx 1.6476\pi$. At this parameter value, the exchange part of the Hamiltonian ${\cal H}_{\rm ex}$ vanishes. Consequently, the Hamiltonian can be written as the sum of a double–spin-flip term ${\cal H}_{\rm dsf}$ and an Ising term ${\cal H}_{\rm Ising}$, each of which is exactly solvable. Furthermore, their eigenstates are orthogonal because they do not share common spin configurations. Consequently, the ground state is two-fold degenerate with energy $E_0 = -N\cos\phi$ at this critical point, implying a first-order transition between the N\'eel-$z$ and LLRR2 phases at $\phi = \tan^{-1}(-2)$. For an infinitesimal increase in $\phi$ from $\tan^{-1}(-2)$, the ground state of ${\cal H}_{\rm Ising}$ becomes the ground state of the full Hamiltonian. As a result, the N\'eel order parameter satisfies $\mathcal{O}_{\mathrm{N\acute{e}el\text{-}z}} = 1$ at $\phi = \tan^{-1}(-2)$, as shown in Fig.~\ref{fig:Dz0c00}(b). Conversely, for an infinitesimal deviation towards smaller $\phi$, the LLRR2 configuration becomes fully realized, yielding $\mathcal{O}_{\mathrm{LLRR2}} = 1$. Note that there exists another parameter point, $\phi = \tan^{-1}(-2) \approx 0.6476\pi \ (\mathrm{mod}\ \pi)$, where ${\cal H}_{\rm ex}$ also vanishes. Although this point corresponds to the transition between the LLRR1 and FM-$z$ phases, the eigenstates of ${\cal H}_{\rm dsf}$ and ${\cal H}_{\rm Ising}$ are not orthogonal. Therefore, unlike the previous case, the model cannot be solved analytically at this point.

\subsubsection{Order-parameter behavior for magnetically disordered phases}

Next, we consider five phases that do not display magnetic LRO: the Haldane phase ($-0.10 \lesssim \phi/\pi < 0.41$), the N\'eel-$xy$ phase ($0.41 \lesssim \phi/\pi < 0.47$), the AFM-KSL phase ($0.47 \lesssim \phi/\pi < 0.53$), the LLRR1 phase ($0.53 \lesssim \phi/\pi < 0.65$), and the FM-KSL phase ($1.45 \lesssim \phi/\pi < 1.54$).

{\it Haldane phase.}---The Haldane phase is characterized by a finite string order parameter $\mathcal{O}_{\mathrm{str}}$. Although this order reflects a hidden AFM structure, it does not correspond to conventional magnetic LRO. At $\phi = 0$, the system is equivalent to the spin-1 AFM Heisenberg chain, and $\mathcal{O}_{\mathrm{str}}$ reaches its maximal value of 0.3711. This value decreases gradually as $\phi$ moves away from zero. Toward the boundary with the N\'eel-$xy$ phase at $\phi/\pi = 0.41$, $\mathcal{O}_{\mathrm{str}}$ decays continuously in a manner consistent with a Berezinskii--Kosterlitz--Thouless transition. If the N\'eel-$xy$ phase is interpreted as an XY-like TLL, providing a power-law decay of the spin--spin correlations, this behavior is consistent with previous studies for spin-$\frac{1}{2}$ KH chain~\cite{Clio2018}. Since the Haldane phase is a gapped VBS state, the spin--spin correlations decay exponentially. A detailed analysis of this behavior is presented in Appendix~\ref{App:correlations}.

{\it KSL and adjacent phases.}---The AFM-KSL and FM-KSL phases are identified by a nonzero dimer order parameter. In contrast, the N\'eel-$xy$ and LLRR1 phases exhibit short-range spin patterns [Figs.~\ref{fig:states}(b) and \ref{fig:states}(c), respectively], but their associated order parameters $\mathcal{O}_{\mathrm{N\acute{e}el\text{-}xy}}$ and $\mathcal{O}_{\mathrm{LLRR1}}$ vanish in the thermodynamic limit [see Appendix~\ref{App:scaling}]. This indicates that they correspond to disordered or critical phases, which can be classified as TLLs. Among the disordered phases, the Haldane--N\'eel-$xy$, N\'eel-$xy$--AFM-KSL, and AFM-KSL--LLRR1 transitions are all continuous. Consequently, as shown in Fig.~\ref{fig:Dz0c00}(a), these transitions cannot be detected using the level-crossing method based on the second derivative of the ground-state energy.

{\it Néel-$xy$ phase.}---In the Néel-$xy$ phase, spins form an AFM pattern in the $xy$-plane while preserving continuous spin-rotational symmetry. Due to the absence of explicit symmetry breaking in the $xy$-plane, the system cannot exhibit true LRO in 1D. Nevertheless, quasi-long-range correlations can be captured via the spin structure factor $S^{\alpha\alpha}(q)$ with $\alpha = x$ or $y$, showing a peak at $q = \pi$. As shown in Appendix~\ref{App:correlations}, the spin--spin correlation exhibits a power-law decay as a characteristic of the TLL.

{\it LLRR1 phase.}---While both LLRR1 and LLRR2 phases show similar local spin patterns, their long-distance behavior differs significantly. The LLRR2 phase exhibits true long-range order in this order parameter, remaining finite in the thermodynamic limit. In contrast, the order parameter in the LLRR1 phase decays with increasing system size, indicating the absence of long-range order. This distinction suggests that LLRR2 corresponds to a symmetry-breaking phase, whereas LLRR1 should be regarded as a disordered or critical phase with strong short-range modulation.

{\it Differences from previous work.}---The main differences between our results and those of the previous study~\cite{You2022} are summarized as follows. First, at $D_z = 0$, Ref.~\cite{You2022} identified most of the region with $J > 0$ ($-0.5 < \phi/\pi < 0.5$), specifically $-0.35 \lesssim \phi/\pi \lesssim 0.47$, as belonging to the Haldane phase. In contrast, our analysis reveals that the interval $-0.35 \lesssim \phi/\pi \lesssim -0.10$ should be classified as the N\'eel-$xy$ phase, while $0.41 \lesssim \phi/\pi \lesssim 0.47$ corresponds to the LLRR1 phase. This finding indicates that the Haldane phase is fragile against Kitaev-type anisotropy, particularly for $D_z < 0$. Furthermore, although the LLRR1 and LLRR2 phases exhibit very similar four-site periodic spin patterns, we find that the LLRR2 phase possesses magnetic LRO, whereas the LLRR1 phase shows only quasi-long-range correlations without true LRO. This distinction was not captured in Ref.~\cite{You2022}.

\subsection{$D_z$-dependence of phases}

\subsubsection{Haldane phase}

\begin{figure}[thb]
	\centering
	\includegraphics[width=0.60\linewidth]{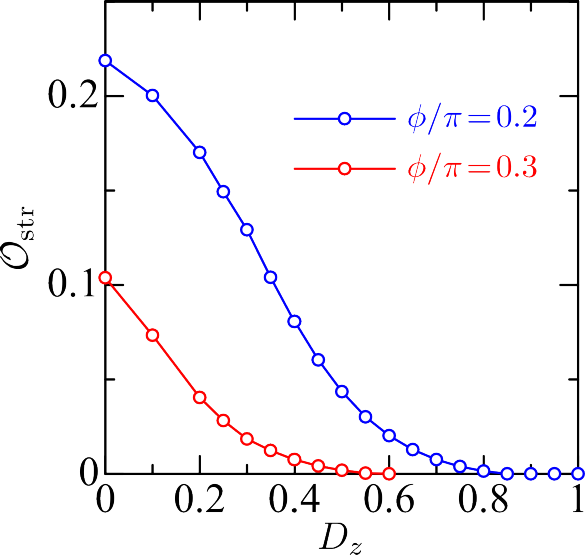}
	\caption{$D_z$ dependence of the string order parameter for (a) $\phi/\pi = 0.2$ and (b) $\phi/\pi = 0.3$.}
	\label{fig:strOP_Dzdep}
\end{figure}

Figure~\ref{fig:strOP_Dzdep} shows the $D_z$ dependence of the string order parameter in the thermodynamic limit for $\phi/\pi=0.2$ and $0.3$. Since the Haldane state is a gapped VBS phase, it is favored by nearly isotropic spin interactions that stabilize effective singlet formation between fractionalized spin-$1/2$ degrees of freedom. Consequently, deviations from the isotropic limit---either by introducing the bond-dependent Kitaev term (finite $\phi$) or by applying a single-ion anisotropy $D_z$---are expected to weaken the Haldane phase. Indeed, $\mathcal{O}_{\mathrm{str}}$ at $\phi/\pi=0.3$ is smaller than that at $\phi/\pi=0.2$, and in both cases it decreases monotonically with increasing $D_z$. The string order parameter smoothly approaches zero upon crossing into the N\'eel-$xy$ phase, consistent with a continuous transition. This trend is analogous to that observed in
Fig.~\ref{fig:Dz0c00}(b) when $\phi/\pi$ is increased at fixed $D_z=0$.

For sufficiently large positive $D_z$, the system is expected to crossover into the so-called large-$D_z$ phase~\cite{Chen2003}, in which the spins are predominantly in the $S^z=0$ state (a trivial product state in the $D_z\!\to\!+\infty$ limit).

\subsubsection{N\'eel-$z$ phase}

\begin{figure}[thb]
	\centering
	\includegraphics[width=0.95\linewidth]{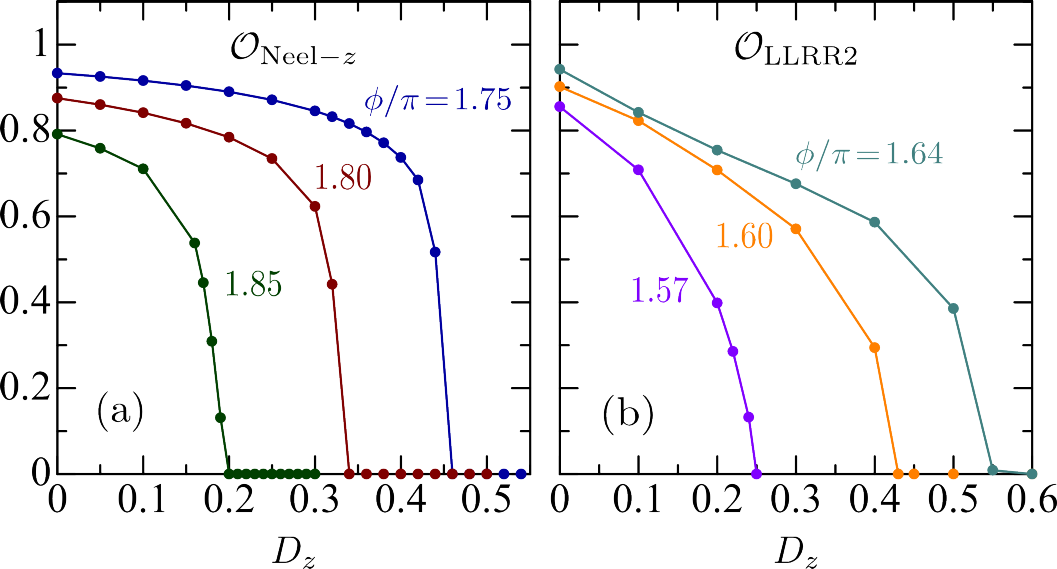}
	\caption{$D_z$ dependence of the N\'eel-$z$ order parameter $\mathcal{O}_{\mathrm{N\acute{e}el\text{-}z}}$ for $\phi/\pi = 1.75$, $1.80$, and $1.85$.}
	\label{fig:LLRR2NeelzOP_Dzdep}
\end{figure}

We now turn to the $D_z$ dependence of the N\'eel-$z$ phase. For $D_z > 0$, the $|\uparrow\rangle$ and $|\downarrow\rangle$ components of the spin-1 basis become progressively less populated compared with the $|0\rangle$ component. As $D_z$ increases, the energy gain from forming a N\'eel state along the $z$ axis is therefore reduced. Figure~\ref{fig:LLRR2NeelzOP_Dzdep} shows $\mathcal{O}_{\mathrm{N\acute{e}el\text{-}z}}$ as a function of $D_z$ for $\phi/\pi = 1.75$, $1.80$, and $1.85$. As $\phi/\pi$ increases, the critical value of $D_z$ decreases, as expected. Near $D_z = 0$, the initial rate of decrease of $\mathcal{O}_{\mathrm{N\acute{e}el\text{-}z}}$ is slightly larger for larger $\phi/\pi$, reflecting the reduced stability of the N\'eel-$z$ order. However, $\mathcal{O}_{\mathrm{N\acute{e}el\text{-}z}}$ drops sharply to zero only when approaching the phase boundary with the Haldane phase.

Although we do not show the data here, for negative $D_z$ the $|\uparrow\rangle$ and $|\downarrow\rangle$ components become more populated, and the spin-1 degrees of freedom effectively reduce to an Ising-like form. In the large negative-$D_z$ limit, the Hamiltonian essentially reduces to ${\cal H}_{\rm Ising}$. As a result, the entire parameter regions $0 < \phi/\pi < 0.5$ and $1.5 < \phi/\pi < 2.0$ are occupied by the N\'eel-$z$ phase in the ground-state phase diagram.

\subsubsection{LLRR2 phase}

The stability of the LLRR2 order is similar to that of the N\'eel-$z$ phase. As discussed above for $D_z=0$ case, the order parameter $\mathcal{O}_{\mathrm{LLRR2}}$ reaches its maximum value of 1 at $\phi = \tan^{-1}(-2)$ and decreases as $\phi/\pi$ is reduced, dropping sharply to zero near the boundary with the FM-KSL phase at $\phi/\pi \approx 1.54$. Fig~\ref{fig:LLRR2NeelzOP_Dzdep}(b) shows $\mathcal{O}_{\mathrm{LLRR2}}$ as a function of $D_z$ for $\phi/\pi = 1.57$, $1.60$, and $1.64$. Compared with $\mathcal{O}_{\mathrm{N\acute{e}el\text{-}z}}$, the stronger quantum fluctuations in the LLRR2 phase lead to a more rapid suppression of $\mathcal{O}_{\mathrm{LLRR2}}$ as $D_z$ increases. The critical value of $D_z$ at which $\mathcal{O}_{\mathrm{LLRR2}}$ vanishes increases with $\phi/\pi$, reflecting the higher intrinsic stability of the LLRR2 order for larger $\phi/\pi$. Importantly, even after $\mathcal{O}_{\mathrm{LLRR2}}$ becomes zero, the short-range four-site periodic spin structure remains intact. This indicates that the transition driven by $D_z$ can be interpreted as a change from LLRR2 LRO to LLRR2 SRO.

\subsubsection{Ferromagnetic states}

\begin{figure}[t]
	\centering
	\includegraphics[width=0.9\linewidth]{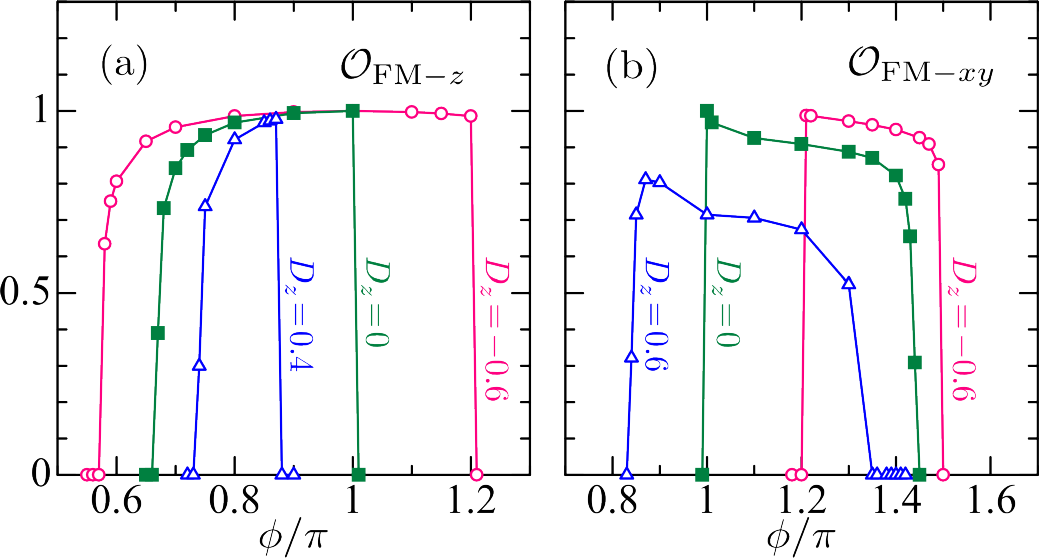}
	\caption{$D_z$ dependence of the ferromagnetic order parameters. 
		(a) $\mathcal{O}_{\mathrm{FM\text{-}z}}$ as a function of $\phi/\pi$ for $D_z = -0.6$, $0$, and $0.4$. 
		(b) $\mathcal{O}_{\mathrm{FM\text{-}xy}}$ as a function of $\phi/\pi$ for $D_z = -0.6$, $0$, and $0.6$.}
	\label{fig:FMOP_Dzdep}
\end{figure}

As the nature of magnetization axes differ between FM-$z$ and FM-$xy$ phases, the $D_z$ dependence of these phases also shows distinct behaviors. Fig~\ref{fig:FMOP_Dzdep}(a) depicts $\mathcal{O}_{\mathrm{FM\text{-}z}}$ as a function of $\phi/\pi$ for $D_z = -0.6$, $0.0$, and $0.4$. For $D_z > 0$, spin polarization along the $z$ axis becomes energetically unfavorable. Consequently, the FM-$z$ region for $D_z = 0.4$ is significantly reduced compared to that of $D_z = 0$. In contrast, for $D_z < 0$, polarization along the $z$ axis yields an energy gain, and the FM-$z$ region for $D_z = -0.6$ is substantially enhanced relative to that at $D_z = 0$. In the limit $D_z \to -\infty$, 
the Hamiltonian effectively reduces to ${\cal H} = {\cal H}_{\rm Ising} = J S_i^z S_{i+1}^z$. As a result, the entire region with negative $J$ ($0.5 < \phi/\pi < 1.5$) becomes FM ordered along the $z$ direction.

Let us now consider the FM-$xy$ phase. Figure~\ref{fig:FMOP_Dzdep}(b) shows $\mathcal{O}_{\mathrm{FM\text{-}xy}}$ as a function of $\phi/\pi$ for $D_z = -0.6$, $0.0$, and $0.6$. The overall extent of the FM-$xy$ region for $D_z = 0.6$ is almost unchanged from that at $D_z = 0.0$, but the magnitude of $\mathcal{O}_{\mathrm{FM\text{-}xy}}$ itself decreases. Although the magnetization in the FM-$xy$ phase does not point along the $z$ axis, its wave function is composed of linear combinations of $|\uparrow\rangle$ and $|\downarrow\rangle$. Therefore, as the weight of the $|0\rangle$ component increases with $D_z$, the stability of the FM-$xy$ order is reduced. For sufficiently large positive $D_z$, the system is expected to evolve into a paramagnetic state with short-range FM-$xy$ correlations. For $D_z = -0.6$, the maximum value of $\mathcal{O}_{\mathrm{FM\text{-}xy}}$ remains close to unity, but the FM-$xy$ region is significantly reduced. This reduction originates not from the destabilization of the FM-$xy$ order itself, but rather from the expansion of the FM-$z$ phase into the $\phi/\pi > 1$ region. In all cases, the transition between the FM-$z$ and FM-$xy$ phases is of the first order.

\subsubsection{Kitaev spin-liquid states}

\begin{figure}[t]
	\centering
	\includegraphics[width=1.0\linewidth]{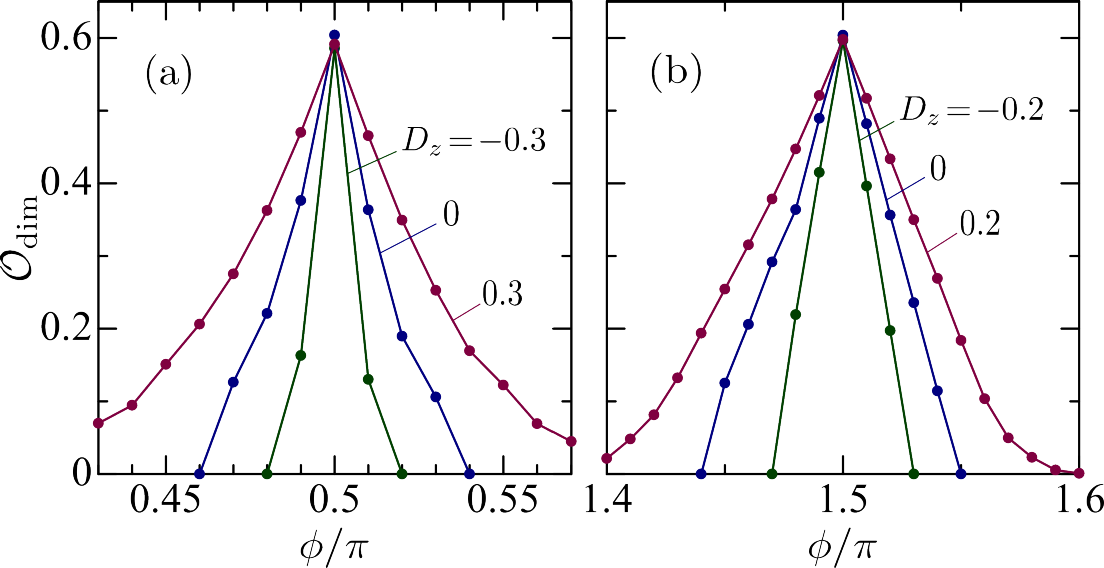}
	\caption{$D_z$ dependence of the dimer order parameter near (a) the AFM-KSL region ($\phi/\pi = 0.5$) and (b) the FM-KSL region ($\phi/\pi = 1.5$).}
	\label{fig:dimer_Dzdep}
\end{figure}

Figure~\ref{fig:dimer_Dzdep} shows the dimer order parameter as a function of $\phi/\pi$ for several values of $D_z$ near (a) the AFM-KSL region ($\phi/\pi \approx 0.5$) and (b) the FM-KSL region ($\phi/\pi \approx 1.5$). Both KSL regions exhibit qualitatively similar $D_z$ dependence-- expands (shrinks) as $D_z$ is increased (decreased) from $D_z=0$. 
This trend is consistent with the previous study~\cite{Zhang2023}. The underlying reason is that the positive $D_z$ suppresses all types of magnetic order, while the negative $D_z$ strongly enhances the N\'eel-$z$ and FM-$z$ phases. For $D_z > 0$, however, a crossover likely occurs between the KSL and paramagnetic states. Therefore, $D_z$ does not directly affect the intrinsic stability of the KSL itself but rather controls the relative stability of the surrounding magnetically ordered phases. This interpretation is further supported by the observation that the magnitude of $\mathcal{O}_{\mathrm{dimer}}$ at $\phi/\pi = 0.5$ and $1.5$ remains nearly unchanged with varying $D_z$.

\section{Discussion}\label{sec:discussion}

\begin{figure}[t]
	\centering
	\includegraphics[width=1.00\linewidth]{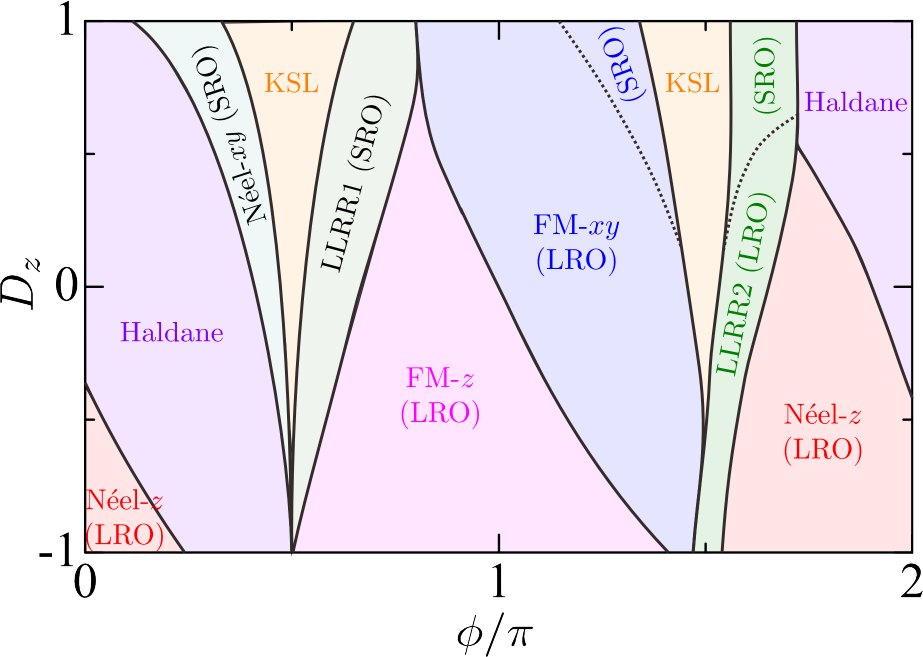}
	\caption{Ground-state phase diagram in the $\phi$--$D_z$ plane.}
	\label{fig:PD_OP}
\end{figure}

All the above discussions/explanations regarding various phases have been summerized to map out the ground-state phase diagram in $\phi/\pi-D_z$ plane as shown in Fig.~\ref{fig:PD_OP}. In general, as $D_z$ increases to positive values, the KSL regions expand while the magnetic orders are suppressed. In contrast, when $D_z$ becomes more negative, the FM-$z$ and N\'eel-$z$ regions are enlarged significantly.

Let us now compare our results with those of the spin-1/2 KH chain~\cite{Clio2018,Yang2020,Yang2025}. Since the $D_z$ term has no physical meaning for spin-1/2, we focus on the case of $D_z = 0$. Although there are quantitative differences in the transition points, the N\'eel-$xy$, LLRR1, FM-$z$, FM-$xy$, LLRR2, and N\'eel-$z$ phases exhibit essentially the same characteristics in both spin-1 and spin-1/2 systems. The major qualitative difference emerges around the Heisenberg limit ($K = 0$), where the TLL phase in the spin-1/2 system is replaced by the Haldane phase in the spin-1 case. Moreover, we have clarified that the AFM-KSL and FM-KSL phases in the spin-1 KH chain occupy finite parameter regions, whereas in the spin-1/2 case it remains unclear whether the KSL exists only at the two isolated points $\phi/\pi = 0.5$ and $1.5$ or extends over finite ranges.

It is instructive to compare our phase diagram with that reported in Ref.~\cite{Zhang2023}, which investigated the effect of $D_z$ in the spin-1 KH chain in the vicinity of the AFM-KSL regime (i.e., within a restricted range of $J$ and $K>0$). While the overall connectivity of the robust magnetically ordered phases is consistent, the main differences appear in the magnetically disordered/critical regimes, where phase identification is more sensitive to the diagnostics and to thermodynamic-limit extrapolations. In our phase diagram, the most notable discrepancies are: (i) between the Haldane and AFM-KSL regimes we find a sizeable intervening N\'eel-$xy$ region; (ii) near the Heisenberg limit ($\phi\simeq 0$), the Haldane phase terminates already at relatively small negative $D_z$ and gives way to the N\'eel-$z$ phase; and (iii) the LLRR1 regime does not develop true LRO, but remains magnetically SRO. Regarding point (ii), we find that at $\phi=0$ the transition to the N\'eel-$z$ phase occurs at $D_z\simeq -0.5$, which is in good agreement with the estimate reported in Ref.~\cite{Chen2003}. A direct graphical comparison over the full $\phi$--$D_z$ plane, including a representation in the parametrization of Ref.~\cite{Zhang2023}, is provided in Appendix~\ref{App:cf_PD} (Figs.~\ref{fig:PD_cf_1} and \ref{fig:PD_cf_2}).

It would be also interesting to compare our results with those for the 2D spin-1 KH model on the honeycomb lattice~\cite{Bradley2022,Ayushi2024}, where the effect of $D_z$ has also been examined. In 2D case, a large negative $D_z$ enhances Ising anisotropy and consequently stabilizes states corresponding to the N\'eel-$z$ and FM-$z$ phases, similar to the 1D case. However, for large positive $D_z$, the KSL region becomes narrower rather than broader. One of the primary reasons for this contrast is that in the honeycomb lattice, positive $D_z$ promotes the formation of magnetic vortex states with enlarged magnetic unit cells, which tend to invade and destabilize the KSL region.

\section{Summary}\label{sec:summary}

Using the DMRG method, we have determined the ground-state phase diagram of the spin-1 KH chain with uniaxial SIA $D_z$ by combining energy-curvature diagnostics on periodic clusters with order-parameter and correlation analyses on larger open chain systems. The resulting $\phi$--$D_z$ phase diagram hosts magnetically ordered FM-$z$, FM-$xy$, LLRR2, and N\'eel-$z$ phases; magnetically disordered/critical N\'eel-$xy$ and LLRR1 regimes; two KSL (AFM- and FM-KSL) regions; and a topological Haldane phase near the Heisenberg limit. Our calculations indicate that the AFM- and FM-KSL regimes occupy finite parameter regions in the spin-1 model. We also clarify that LLRR1 and LLRR2, despite exhibiting similar four-site spin patterns, differ fundamentally in that only LLRR2 develops true LRO.

SIA provides an efficient knob to balance frustration, quantum fluctuations, and topology: positive $D_z$ suppresses magnetic order and broadens the KSL and other magnetically disordered sectors, whereas negative $D_z$ stabilizes Ising-like N\'eel-$z$ and FM-$z$ orders. At $D_z=0$, we identify an exactly solvable point at $\phi=\tan^{-1}(-2)$ which enforces a first-order boundary between N\'eel-$z$ and LLRR2 phases. These trends rationalize the evolution of the phase diagram and help explain why the Haldane phase is particularly fragile against Kitaev-type anisotropy for $D_z<0$.

Finally, comparison with the spin-$1/2$ KH chain shows qualitative agreement for most phases but a key replacement of the TLL at $K=0$ by the Haldane phase in the spin-1 case; it also highlights that finite KSL widths are supported in spin-1, while their extent in spin-$1/2$ remains unsettled. In contrast, on the spin-1 honeycomb lattice, large positive $D_z$ favors vortex states with enlarged magnetic unit cells that encroach upon and narrow the KSL sector, underscoring the crucial role of dimensionality and anisotropy in Kitaev-type magnets.

{\it Acknowledgements.---}
We thank Ulrike Nitzsche for technical support. This work is supported by
the SFB 1143 of the Deutsche Forschungsgemeinschaft (DFG) via the projects A05 of the Collaborative Research Center SFB 1143 (project-id 247310070), SRG/2020/001203 of SERB, India, and F.30-528/2020 (BSR) of UGC, India. A part of the computation is performed in Physics Department, Jadavpur University Computer Facility (Sanction No. SR/FST/PS-1/2022/219(C))

\appendix

\section{Finite-size scaling for the order parameters}\label{App:scaling}

\begin{figure}[t]
	\centering
	\includegraphics[width=0.90\linewidth]{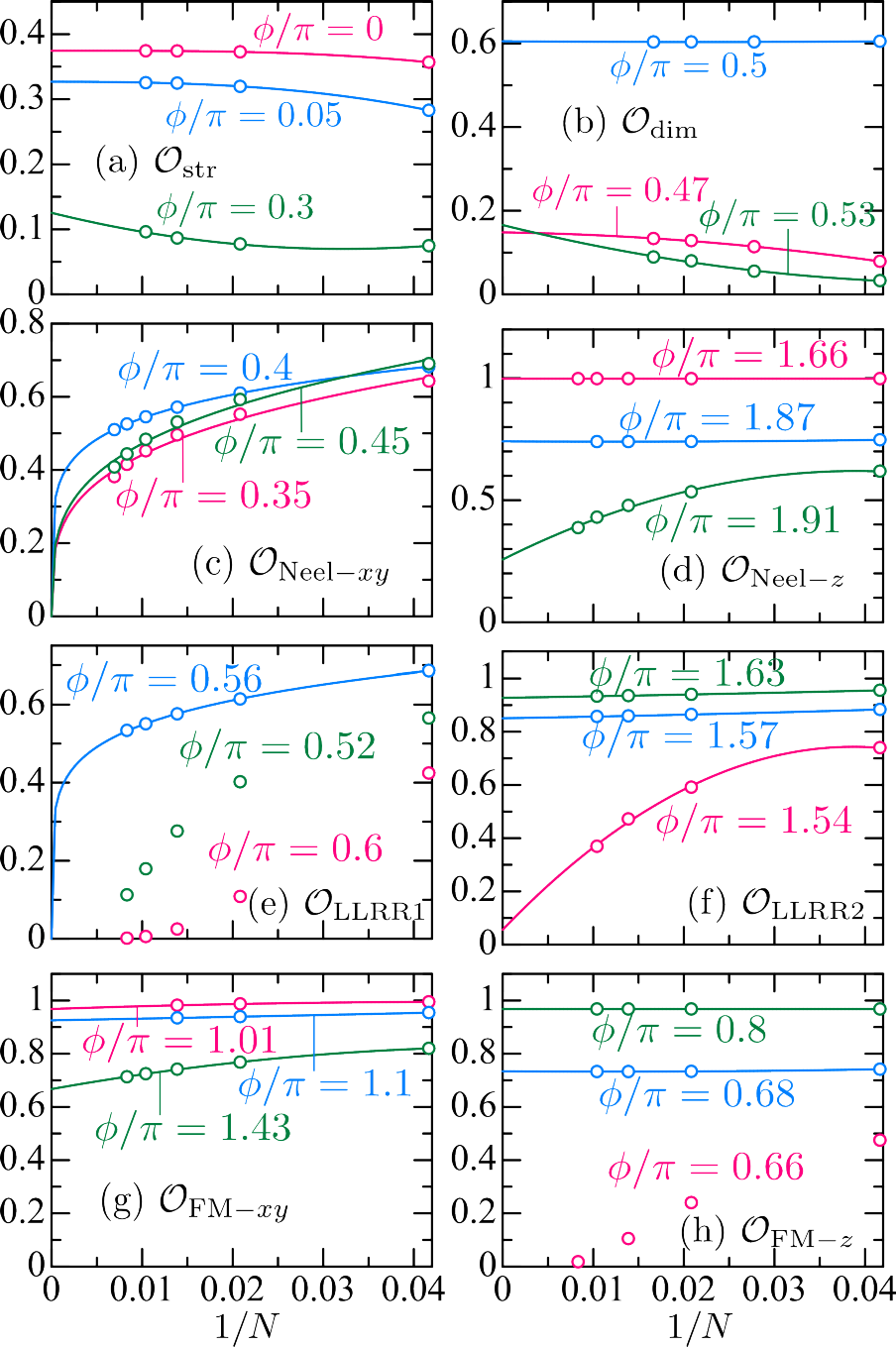}
	\caption{Finite-size scaling of the order parameters for representative parameters at $D_z=0$.}
	\label{fig:scaling}
\end{figure}

In this Appendix, we present representative examples of the finite-size scaling analyses used to obtain the thermodynamic-limit values of the order parameters. Figure~\ref{fig:scaling} shows the size dependence of the order
parameters for typical points in each phase at $D_z=0$, computed on open chains with lengths ranging from $N=24$ up to $N=144$.

To extrapolate ${\cal O}(N)$ to the thermodynamic limit, we employ two types of fitting functions depending on the convergence behavior. For the LLRR1 and N\'eel-$xy$ phases, where the order parameters approach their asymptotic
values relatively slowly with increasing $N$, we use a power-law form,
\begin{equation}
	{\cal O}(N)=\frac{a}{N^{b}}+c.
\end{equation}
For the other phases, the finite-size corrections are well described by a second-order polynomial in $1/L$,
\begin{equation}
	{\cal O}(N)=\frac{a}{N^{2}}+\frac{b}{N}+c.
\end{equation}
In all cases, the extrapolated value $c$ is taken as the order parameter in the thermodynamic limit.

\section{Distinct decay of spin--spin correlations in the Haldane and N\'eel-$xy$ phases}\label{App:correlations}

\begin{figure}[t]
	\centering
	\includegraphics[width=1.00\linewidth]{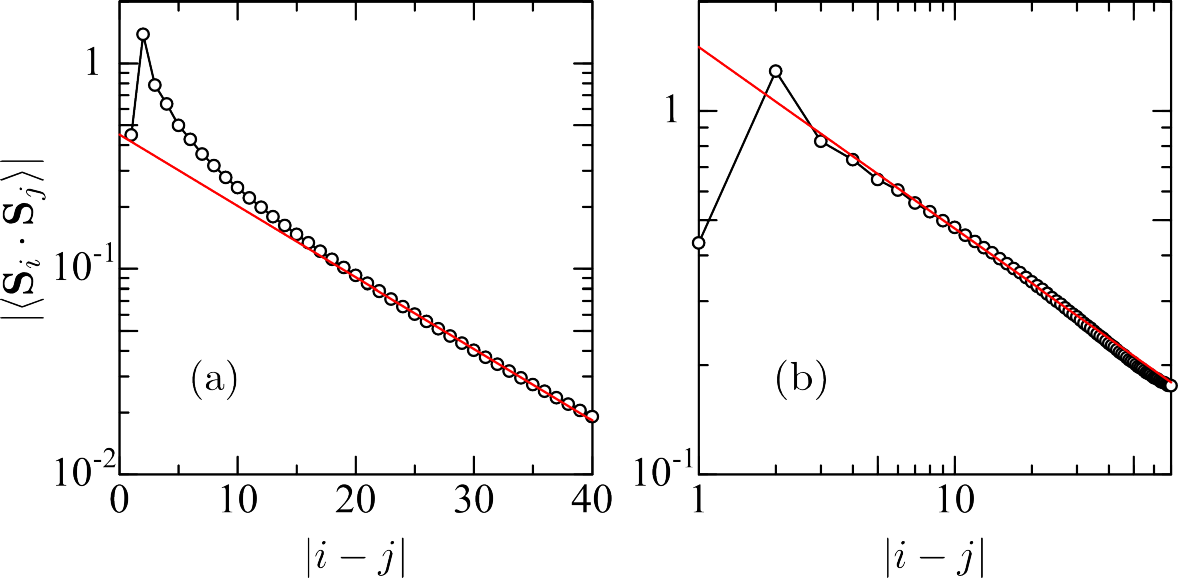}
	\caption{Spin--spin correlation function $|\langle \mathbf{S}_i\cdot \mathbf{S}_j\rangle|$ as a function of the distance $r=|i-j|$ for representative parameters in the (a) Haldane and (b) N\'eel-$xy$ phases. The solid lines show fits to the long-distance behavior using an exponential decay in (a) and a power-law decay in (b).}
	\label{fig:correlations}
\end{figure}

As discussed in the main text, the Haldane phase is a gapped VBS state. Consequently, the spin--spin correlation function decays exponentially with distance. By contrast, the adjacent N\'eel-$xy$ phase is characterized as a TLL, for which spin correlations exhibit a power-law decay. Figure~\ref{fig:correlations} displays $|\langle \mathbf{S}_i\cdot \mathbf{S}_j\rangle|$ as a function of the distance $|i-j|$ for representative points in these two phases. As expected, we clearly observe an exponential decay in the Haldane phase and a power-law decay in the N\'eel-$xy$ phase.

In the N\'eel-$xy$ phase, the long-distance tail is well fitted by
$|\langle \mathbf{S}_i\cdot \mathbf{S}_j\rangle| \propto |i-j|^{-0.5}$. This slow decay is consistent with the slow convergence with system size observed for the order parameter ${\cal O}_{{\rm N\acute eel}\text{-}xy}$ in the finite-size scaling analyses.

\section{Comparison with previous phase diagrams}\label{App:cf_PD}

\begin{figure}[tb]
	\centering
	\includegraphics[width=0.80\linewidth]{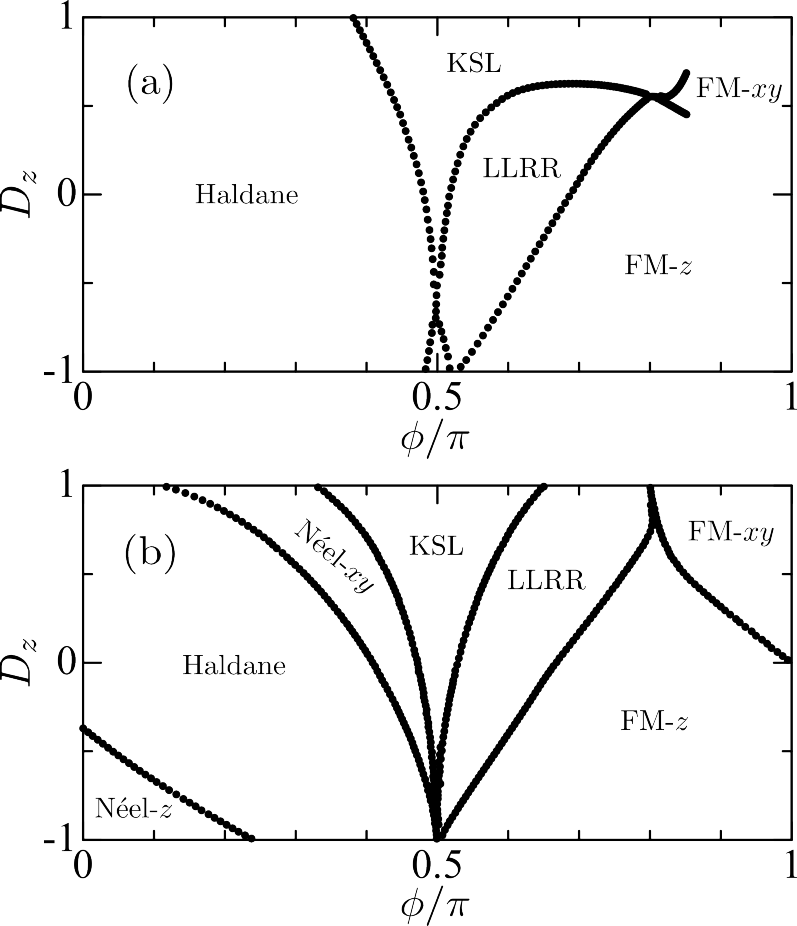}
	\caption{Comparison between our phase diagram and that of Ref.~\cite{Zhang2023} in our parametrization. The two diagrams are shown side by side in the $\phi/\pi$--$D_z$ plane to facilitate a direct visual comparison of the location and connectivity of the phases.}
	
	\label{fig:PD_cf_1}
\end{figure}

\begin{figure}[b]
	\centering
	\includegraphics[width=0.80\linewidth]{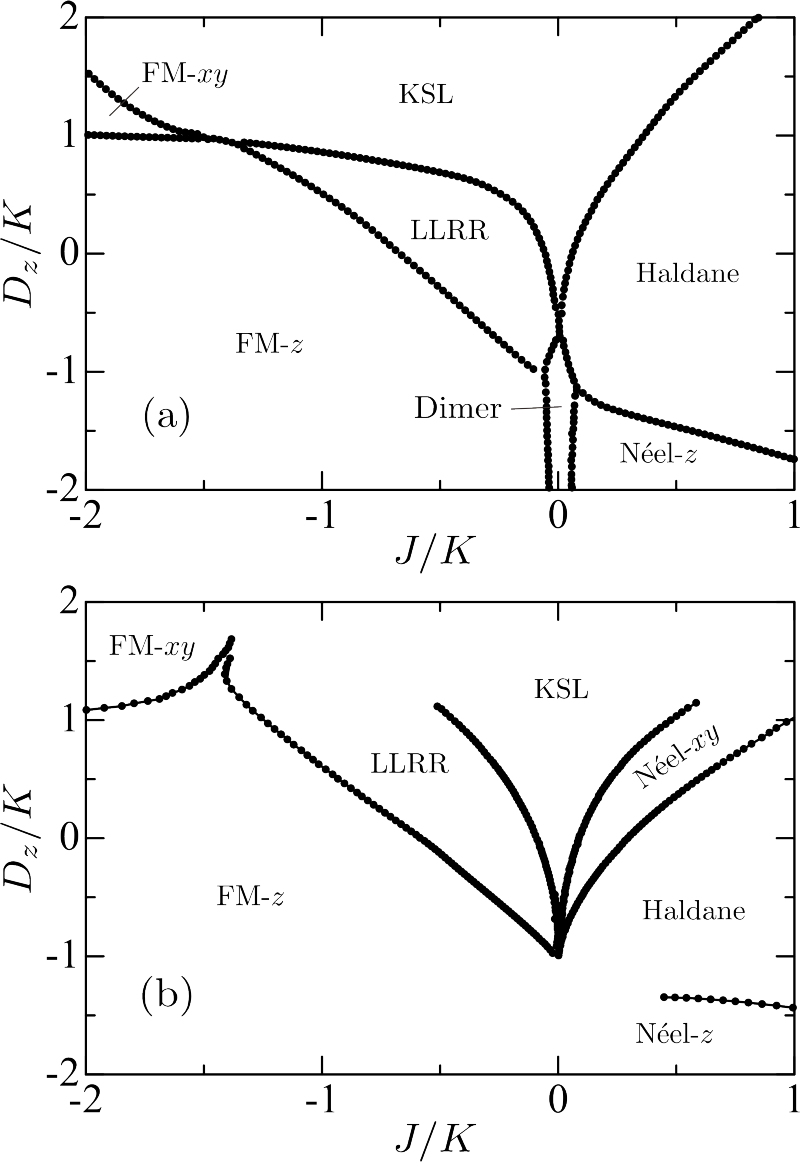}
	\caption{Same comparison as in Fig.~\ref{fig:PD_cf_1}, but displayed in the notation used in Ref.~\cite{Zhang2023}. Here we convert our parameters to the corresponding ratios employed in Ref.~\cite{Zhang2023} (e.g., $J/K$ and $D_z/K$), enabling a one-to-one comparison with their phase boundaries.}
	\label{fig:PD_cf_2}
\end{figure}

For completeness, we compare our phase diagram with that reported in Ref.~\cite{Zhang2023}. Figure~\ref{fig:PD_cf_1} shows a direct side-by-side comparison using our parametrization in the $\phi/\pi$--$D_z$ plane, while Fig.~\ref{fig:PD_cf_2} presents the same comparison after converting to the notation employed in Ref.~\cite{Zhang2023}. Overall, we find good agreement for the main magnetically ordered regions and their overall connectivity. The most notable differences occur in the magnetically disordered/critical regimes, where phase identification is more subtle and sensitive to the diagnostics used. In particular, at $D_z=0$ and for $J>0$, Ref.~\cite{Zhang2023} assigns a broad parameter interval to the Haldane phase, whereas our analysis based on the thermodynamic-limit behavior of order parameters and correlation functions resolves sizeable portions of this region into the N\'eel-$xy$ phase and the LLRR1 regime. Relatedly, we explicitly distinguish LLRR1 and LLRR2: although they exhibit similar local spin patterns, we find that LLRR2 develops true LRO, while LLRR1 does not show a finite order parameter in the thermodynamic limit. These refinements lead to a reduced extent of the Haldane region and modified boundaries near the adjacent critical regimes, while leaving the robust ordered phases essentially unchanged.

\bibliography{S1KHchain_Dz}

\end{document}